\documentclass[aps,pre,twocolumn,superscriptaddress]{revtex4-1}
\pdfoutput=1
\usepackage{amsmath}
\usepackage{subfigure}
\usepackage{color}
\usepackage{graphicx}

\begin{document}

\title{Spatio-Temporal Techniques for User Identification \\ by means of GPS Mobility Data}

\author{Luca Rossi}
\affiliation{School of Computer Science, University of Birmingham, UK}

\author{James Walker}
\affiliation{School of Computer Science, University of Birmingham, UK}

\author{Mirco Musolesi}
\affiliation{School of Computer Science, University of Birmingham, UK}
\affiliation{Department of Geography, University College London, UK}

\begin{abstract}
One of the greatest concerns related to the popularity of GPS-enabled devices and applications is the increasing availability of the personal location information generated by them and shared with application and service providers. Moreover, people tend to have regular routines and be characterized by a set of ``significant places'', thus making it possible to identify a user from his/her mobility data.

In this paper we present a series of techniques for identifying individuals from their GPS movements. More specifically, we study the uniqueness of GPS information for three popular datasets, and we provide a detailed analysis of the discriminatory power of speed, direction and distance of travel. Most importantly, we present a simple yet effective technique for the identification of users from location information that are not included in the original dataset used for training, thus raising important privacy concerns for the management of location datasets.
\end{abstract}

\maketitle

\section{Introduction}
Current and past location information can be considered as the most sensitive data for an individual~\cite{bohn2005social,krumm2009survey}. This is particularly true when entire \textit{trajectories} of individuals are collected and stored by applications and service providers.
Indeed, companies, such as telecommunication operators and service providers, and governmental organizations have access to large collections of person and communication data, which may be used for maintaining and managing communications services, security and surveillance: these include person location data, which can be collected from GPS devices, cellular phone usage and WiFi hotspots. 

In particular, with the increasing availability and popularity of embedded GPS receivers into personal devices and the ability to locate cellular phone users from their interactions with network antennas~\cite{hightower2001location}, new opportunities arise for gaining knowledge about person movement behavior. An increasing number of researchers has been investigating new ways to mine this wealth of location-based data. Examples include the prediction of the future location of a person~\cite{ashbrook2003using}, their mode of transport~\cite{zheng2008understanding} and the identification of individuals from a sample of their location data~\cite{de2013unique}. In~\cite{gonzalez2008understanding} it was shown that there is a high degree of temporal and spatial regularity in human trajectories: people are more likely to visit an area if they have been frequently visited it in the past. Moreover, the time a person returns to a location is very likely to be close to that of his/her previous visits. Thus, given a geographic trajectory, i.e., a collection of chronologically ordered visited locations, a potential attacker can discover a considerable amount of information about that person, such as their home, place of work, interactions with other people and visits to sensitive locations.

The focus of this work is on location based fingerprinting: the aim is to identify individuals from their movement behavior. As with identifying individuals by the ridges on their finger, the ability to identify them by their mobility traces depends on the \textit{uniqueness} of the mobility data associated with them. By uniqueness here we mean the extent to which a recorded location in a dataset is shared among different individuals, i.e., the less shared a location is, the more unique it is. Also, as with traditional fingerprinting, some information about the person to be identified needs to have been previously recorded. A recent contribution in this sense is represented by the work of de Montjoye et al.~\cite{de2013unique}, where the authors are able to identify users from a small subset of their location records taken from mobile phone service antennas. We would like to underline a major difference between this work and that by de Montjoye et al., as in theirs the training set also includes the points used for the testing and the mobility traces are extracted from mobile operators' call data records, instead of exact GPS points.%
%
%
%
%

In this paper, to the best of our knowledge, we present the first evaluation of the uniqueness of GPS data traces and we show that, with the high precision of GPS, a small number of mobility points, even not present in the given mobility databases used for classification, is sufficient to accurately identify individuals. More specifically, the contribution of our work is threefold: Firstly, we show that it is possible to identify individuals with great accuracy using various types of movement data such as speed, direction and distance of travel recorded by means of GPS devices. This suggests that additional care is necessary when anonymized data, also not containing exact geographic coordinates, are publicly released. Secondly, we provide an extensive evaluation of the uniqueness of GPS mobility traces by means of three real-world datasets, namely CabSpotting~\cite{epfl-mobility}, CenceMe~\cite{cenceme} and GeoLife~\cite{geolife-dataset}. We consider both spatial as well as spatio-temporal information, and we show that, in the datasets being investigated, as little as two points are sufficient to uniquely identify nearly all the users. We also evaluate the impact of the dataset size and the precision of the GPS coordinates on the uniqueness of the data. Our findings show that, in some datasets, it is possible to reduce the average uniqueness by means of spatio-temporal coarsening and achieve a given $k$-anonymity~\cite{sweeney2002achieving,abul2008never}. Finally, we introduce a simple yet effective technique for the identification of users from location information that are not included in the original dataset used for extracting user mobility signatures. We also propose a way to measure the extent to which a dataset can resist to an identification attack based on the techniques proposed in this paper.

The remainder of this article is organized as follows: Section~\ref{datasets} describes the datasets used in this study. Section~\ref{methodology} introduces our framework for the evaluation of the uniqueness of mobility data and the identification of users by means of previously unseen points. Section~\ref{experiments} presents an extensive experimental evaluation on real-world datasets, and we summarize our main findings in Section~\ref{discussion}. Finally, we review the related work in Section~\ref{related} and we conclude the paper summarizing its main contributions in Section~\ref{conclusion}.

\section{Dataset Description}\label{datasets}
In this study we consider three widely used mobility datasets, namely CabSpotting~\cite{epfl-mobility}, CenceMe~\cite{cenceme} and GeoLife~\cite{geolife-dataset}. Note that we will use only the latitude, longitude and timestamp values from these datasets, and discard any other additional information, e.g., altitude. Moreover, the traces in all datasets are anonymized and each mobility trace is given a pseudo-identity.

\textbf{CabSpotting}~\cite{epfl-mobility} is a GPS trace collection of 536 taxi cabs in the San Francisco Bay Area taken over a period of 30 days. The locations are recorded with a spatial and temporal resolution of 5 decimal digits and 1 second, respectively. This dataset has been recently used by Piorkowski et al.~\cite{comsnets09piorkowski} to show that certain macroscopic characteristics specific to clustered mobile wireless networks are prevalent in real mobility traces. 
The inherent characteristics of this dataset, such as common routes of taxis and the fact that the trajectories are spatially constrained to lie on the streets, i.e., the points are less unique, make it particularly challenging and thus of special interest for our study. 

\textbf{CenceMe}~\cite{cenceme} is a dataset of GPS recorded locations with high-level user activity, such as sitting, walking and running, collected by means of mobile phones and involving 20 participants during 2 weeks. The locations are recorded with a spatial and temporal resolution of 6 decimal digits and 1 hour, respectively. The participants are students and staff members of the Departments of Computer Science and Biology at Dartmouth College, with most of the participants activity based in the town of Hanover, in New Hampshire, USA. The dataset was originally collected to study new techniques for the optimization of continuous sensing applications~\cite{musolesi2010supporting}. Despite being a relatively small sized dataset, we decided to include it in this study because of its interesting characteristics. In fact, the locations of the participants during the day are likely to be confined to a limited set of academic buildings and recreational facilities on campus. 

\textbf{GeoLife}~\cite{geolife-dataset} is a dataset of GPS traces of 182 users recorded over a period of five years, from April 2007 to August 2012. The dataset was collected by Microsoft Research Asia and contains information about participants mainly located in Beijing. The locations are recorded with a spatial and temporal resolution of 6 decimal digits and 1 second, respectively. For the purpose of the present study, we limit our analysis to the period from January 2008 to December 2008, thus discarding any user that was not active in this time window. We also exclude from our analysis those participants that are not located in Beijing, yielding a total of 70 users. We stress that this is done to maximize the spatial and temporal overlap of the trajectories by excluding those that are spatially or temporally isolated and restricting the analysis to the period and region of maximum activity: this process increases the complexity of the identification task.

\section{Methodology}\label{methodology}

In this section, we show how to evaluate the uniqueness of GPS mobility traces and we propose a technique for the identification of people by means of mobility data. To this end, we propose to use a distance function between a trajectory and a set of sampled points where the spatial distance between two spatio-temporal points is smoothed according to their temporal difference. 
Note that in this paper we assume that, given a dataset, each person is assigned a single trajectory, where a trajectory is a set of GPS points visited by the individual. More precisely, each GPS point $p$ is a triplet $(lat_p,long_p,time_p)$ defining the spatial and temporal coordinates of $p$.

\subsection{Classification of Previously Seen Points}\label{movement}
We first consider a scenario in which the attacker is given a number of anonymized points sampled from a person's mobility trace and tries to identify which mobility trace these points came from, by comparing the given points to a dataset of mobility traces. This type of attack relies on the underlying uniqueness of a person mobility trace and thus it is considered successful if comparisons reveal the given points can be associated with a small number of person mobility traces. By uniqueness here we mean the extent to which the data is shared among different individuals, i.e., the less shared a location is, the more unique it is. Note that in this scenario the points are not removed from the mobility trace, and, as a result, the design of this classification system is straightforward and computationally inexpensive. In our implementation each point of a user is stored in a hash set, which allows searching in constant time. When given a set of points to classify, we simply identify the number of users which contain all of them.


In addition to this, we also examine the situation in which the attacker has access to alternative movement information. More precisely, we study the uniqueness of information describing either the distance covered, the average speed or the average angle of travel, over a given time window. Recall that GPS coordinates usually consist of pairs of \emph{latitude, longitude} points. In order to compute the distance between two pairs of GPS coordinates we use the well known Haversine formula~\cite{robusto1957cosine}, which gives the shortest distance in kilometers between two locations along the surface of the Earth in a suitable metric\footnote{Our implementation of the Haversine formula uses a general radius of the Earth, and does not truly account for altitude regions such as hills. More accurate distances can be calculated with additional information of the area in which the distance has been calculated.}.
We measure the direction of travel between two points as the initial bearing.
The average direction over a specified time interval is calculated by weighting the direction by the distance traveled in that direction. For example, if a user travels 1 km in the direction $45^\circ$ and 2 km in the direction $90^\circ$, the weighted average direction in this case would be $75.36^\circ$. With the additional information of a timestamp, kilometers per hour speed can be easily calculated as well.

\subsection{Classification of Previously Unseen Points}\label{classification}

In the classification of unseen spatial and spatio-temporal points, an attacker is given a sample of anonymized points $P$ which have been removed from a person's mobility trace $M$, i.e., a set of visited spatio-temporal points. As in the classification of previously seen points, an attack is successful if it associates correctly these points to one or a small number of person mobility traces. Unlike the classification of previously seen points, it is difficult for an attacker to fully validate the correctness of the results from their attack, as the given points may seem to be most similar to one mobility trace, when in reality they belong to another person's mobility trace. This type of attack assumes that there exists a relationship between points in a user mobility trace, i.e., the points being classified should lie spatially and temporally close to the trajectory of the user they belong to. Provided that the spatio-temporal points are sampled densely enough, this is indeed generally true, as researchers have shown that there is a high degree of temporal and spatial regularity in human trajectories~\cite{gonzalez2008understanding}. Note that unlike in the scenario described in the previous subsection, the design of this identification method is more complicated and computationally expensive, due to the need to evaluate the distance function between a given sample set of points and a set of mobility traces. Finally, we stress that in a general scenario the set of points $P$ and the mobility trace $M$ may actually belong to different datasets. In this case, the task of the attacker is that of transferring the identity information from the labeled dataset containing $M$ to the anonymized set of points $P$.

\begin{figure}[t]
\centering
\includegraphics[width=0.8\linewidth]{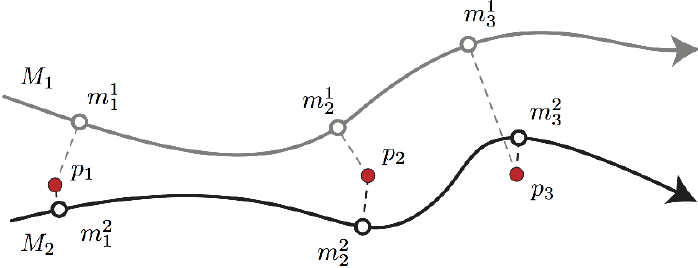}
\caption{Trajectory-based identification. Two traces $M_1$ (grey) and $M_2$ (black) along with a set of three points (red) sampled from $M_2$. These points are classified as belonging to $M_2$ because the average distance to the corresponding nearest points in $M_2$ is lower than the average distance to the nearest points in $M_1$.}
\label{groupdist}
\end{figure} 

In order to evaluate the similarity between sets of spatio-temporal points, we propose to adapt the modified Hausdorff distance~\cite{dubuisson1994modified} to our problem. Recall that the Hausdorff distance between two finite sets of points $A=\lbrace a_1 , \cdots , a_m \rbrace$ and $B = \lbrace b_1, \cdots , b_n \rbrace$ is defined as
\begin{equation}
H(A,B) =  \max(h(A,B),h(B,A))
\end{equation}
where $h(A,B)$ is the directed Hausdorff distance from set $A$ to $B$
\begin{equation}
h(A,B) =  \max_{a \in A} \min_{b \in B} || a-b ||
\end{equation}
and $|| \cdot ||$ denotes the norm on the underlying space. The modified Hausdorff distance is introduced by Dubuisson et al.~\cite{dubuisson1994modified} as
\begin{equation}
h_m(A,B) = \frac{1}{|A|}\sum_{a \in A}\min_{b \in B} || a-b ||.
\end{equation}
where $|A|$ denotes the number of points in $A$.

\begin{figure*}[t!]
\centering
\subfigure[CabSpotting]{\includegraphics[width=0.31\linewidth]{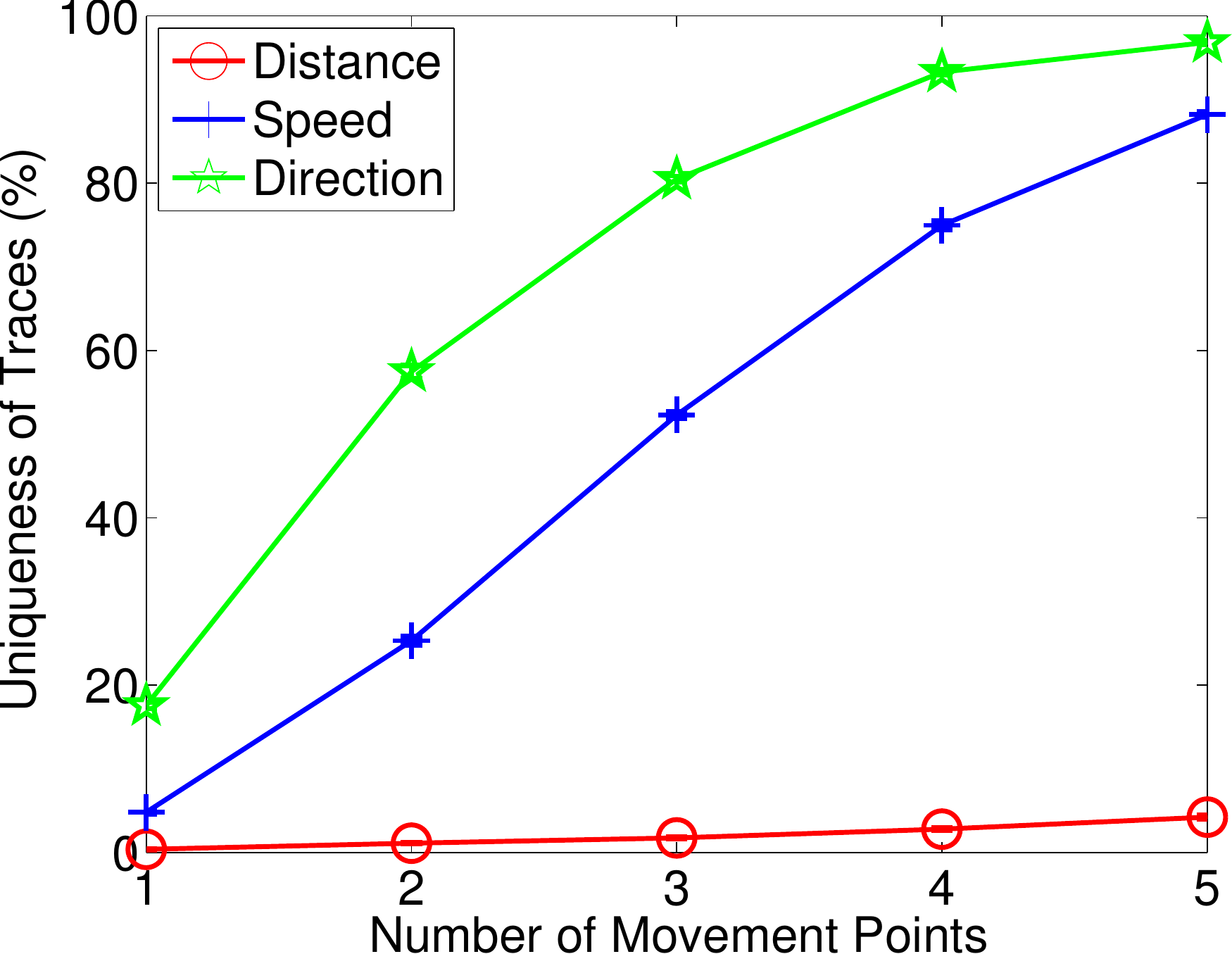}}
\hspace{0.01in}
\subfigure[CenceMe]{\includegraphics[width=0.31\linewidth]{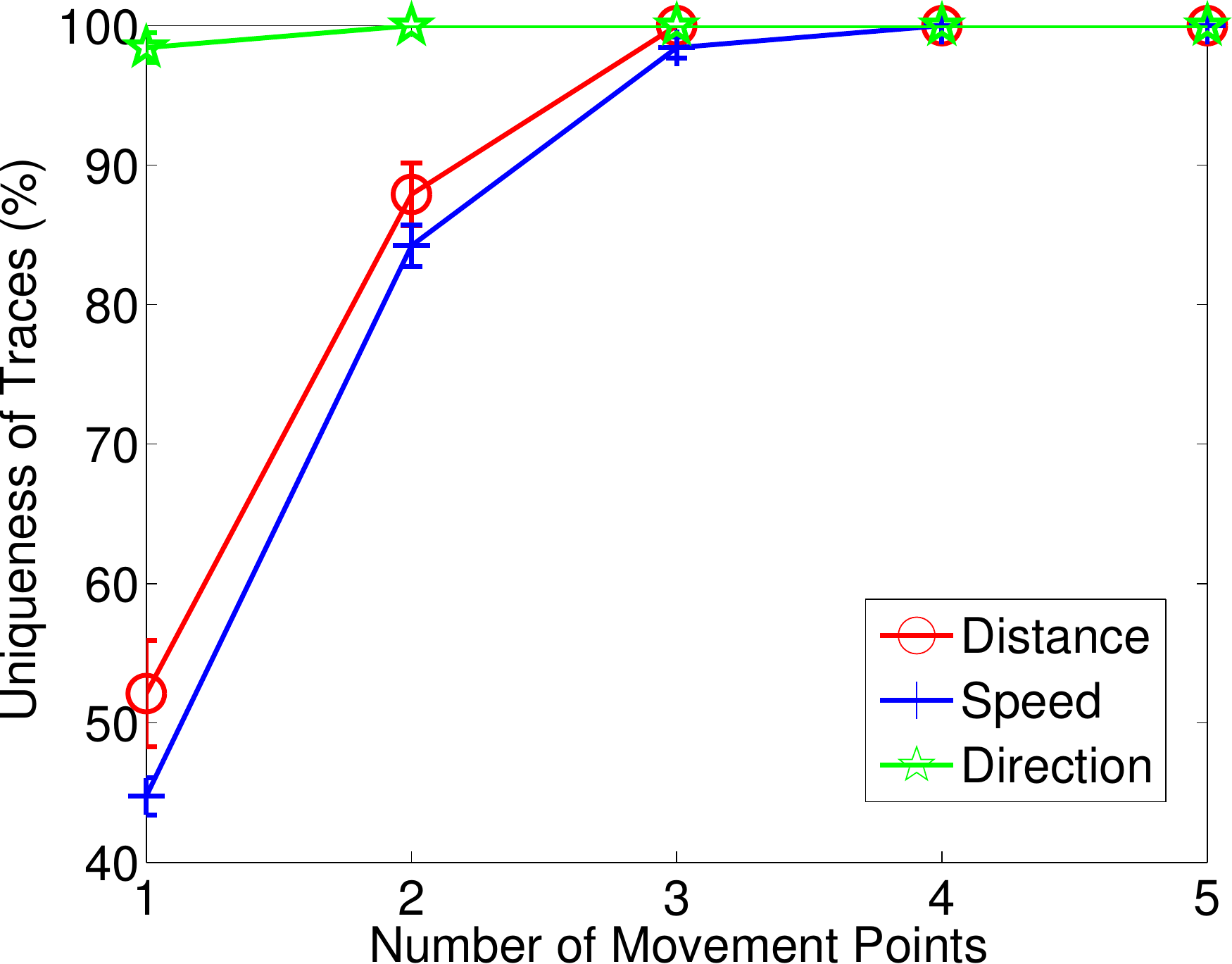}}
\hspace{0.01in}
\subfigure[GeoLife]{\includegraphics[width=0.31\linewidth]{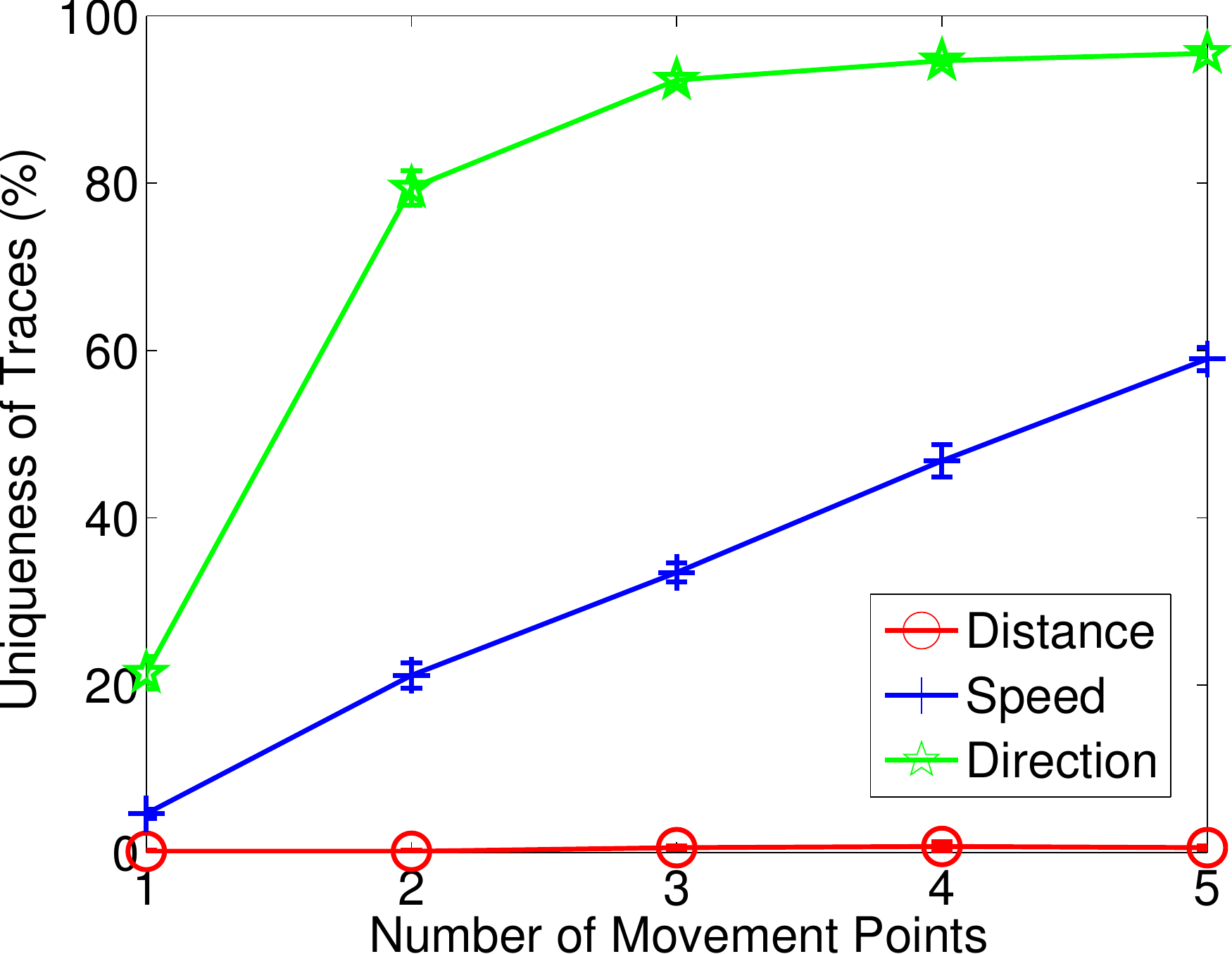}}
\caption{Average uniqueness of movement information. For each dataset, we measure the average uniqueness of the users' movements (y-axis) as we vary the number of movement points considered (x-axis). Specifically, each movement point registers the average distance (red), speed (blue) or direction (green) of travel over a time window of 30 seconds. As expected, the average uniqueness increases as we add more movements points. The effect varies over the three datasets, but in all of them the average direction of travel appears to be the most discriminative feature.}
\label{fig:movement}
\end{figure*}

In order to extend the modified Hausdorff distance to our setting, we start by defining the spatio-temporal distance $d_{st}(p_1,p_2)$ between two points $p_1$ and $p_2$ as
\begin{equation}\label{distance}
d_{st}(p_1,p_2)=d_s(p_1,p_2) e^{\frac{d_t(p_1,p_2)}{\tau}}
\end{equation}  
where $d_s$ denotes the distance computed using the Haversine formula, while $d_t$ denotes the absolute time difference between two points. Here the exponential is used to smooth the distance between two points according to their absolute difference of their timestamps. Note that by setting $\tau \rightarrow \infty$ we ignore the temporal dimension, i.e., the distance between two spatio-temporal points reduces to their Haversine distance. 

With Eq.~\ref{distance} to hand, we can define the distance $d(P,M)$ between a sample set of points $P$ and a mobility trace $M$ as 
\begin{equation}\label{final_distance}
d(P,M) = \frac{1}{|P|} \sum_{p \in P} \min_{m \in M} d_{st}(p,m)
\end{equation}
Fig.~\ref{groupdist} shows the intuition behind the use of this distance function, which can be understood as the average distance to the nearest point in $M$ for every point in $P$. We stress that our distance function is not properly a metric, as it is not symmetric. However, we choose this distance measure for its ease of implementation and its robustness to outliers~\cite{dubuisson1994modified}. As we will show in the experimental part, the set $P$ may contain as little as 1 point, and thus our distance function should be fit to work with a small number of sample points $P$. On the other hand, if we were to take the whole trajectory into account when computing the distance between a single point and $M$, we would inevitably end up overestimating the distance between the point and $M$. Hence, we had to avoid the use of other more popular distance functions such as the classic Hausdorff distance, the Fr{\'e}chet distance~\cite{sack1999handbook}, or the Dynamic Time Warping distance~\cite{berndt1994using}, which are known to be particularly sensitive to outliers~\cite{sack1999handbook,ratanamahatana2004everything,brakatsoulas2005map}. In fact, these metrics either take the maximum over the set of distances between the points in $M$ and $P$, or always try to match $P$ to the whole trajectory $M$. In other words, given the nature of our problem, where we do not consider trajectories as sequences of segments, or paths, but merely as sets of time labeled points, and where the sizes of $M$ and $P$ can be extremely different, these metric are not suitable. On the other hand, the Hausdorff distance, and in particular its modified version, represents a natural choice for our problem.

\section{Experimental Evaluation}\label{experiments}
In this section we perform an extensive experimental evaluation of the techniques introduced in Section~\ref{methodology} on the real-world mobility datasets described in Section~\ref{datasets}.
 
\subsection{Characterization of the Uniqueness of the Mobility Traces}

\begin{figure*}[t!]
\centering
\subfigure[CabSpotting]{\includegraphics[width=0.31\linewidth]{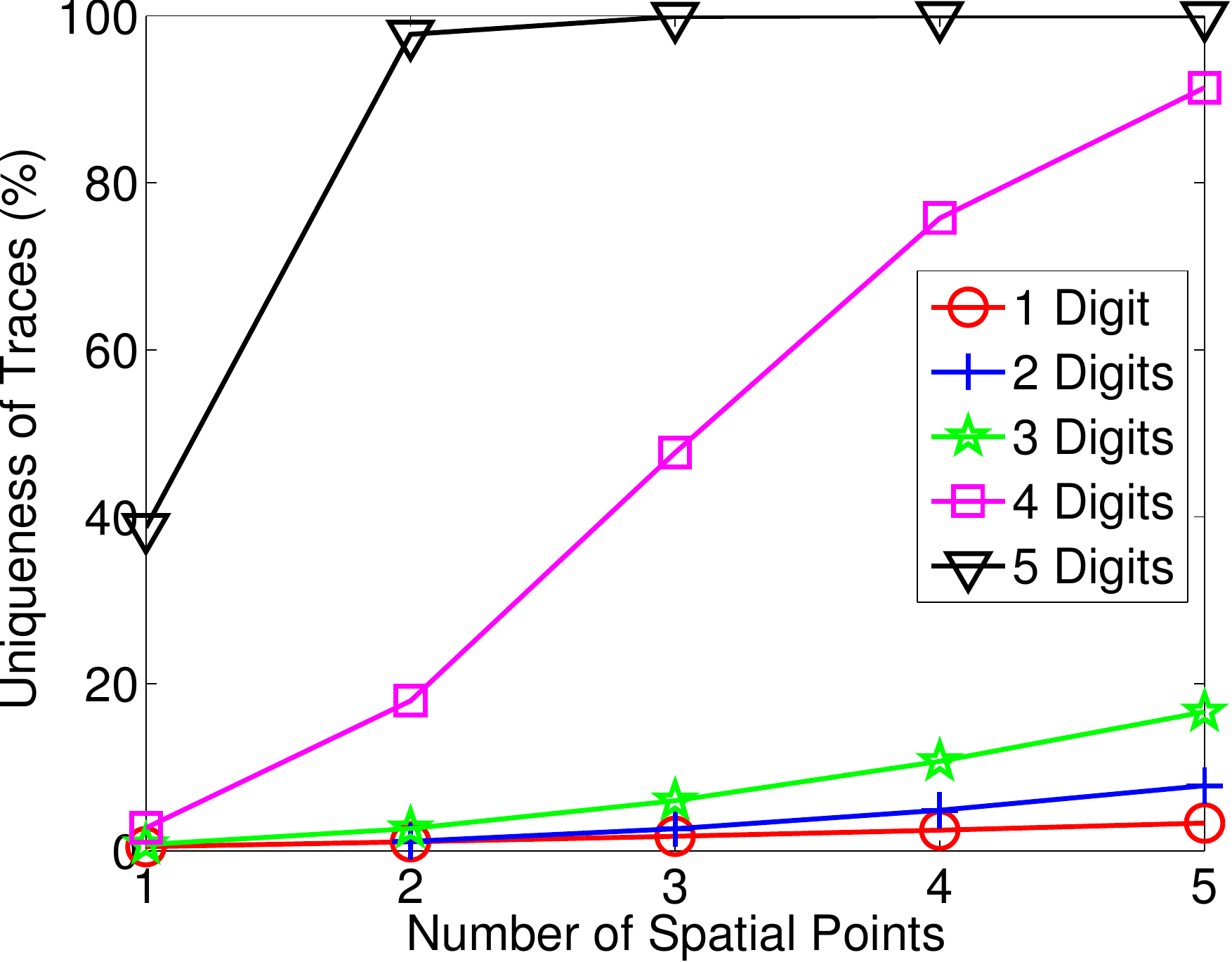}}
\hspace{0.01in}
\subfigure[CenceMe]{\includegraphics[width=0.31\linewidth]{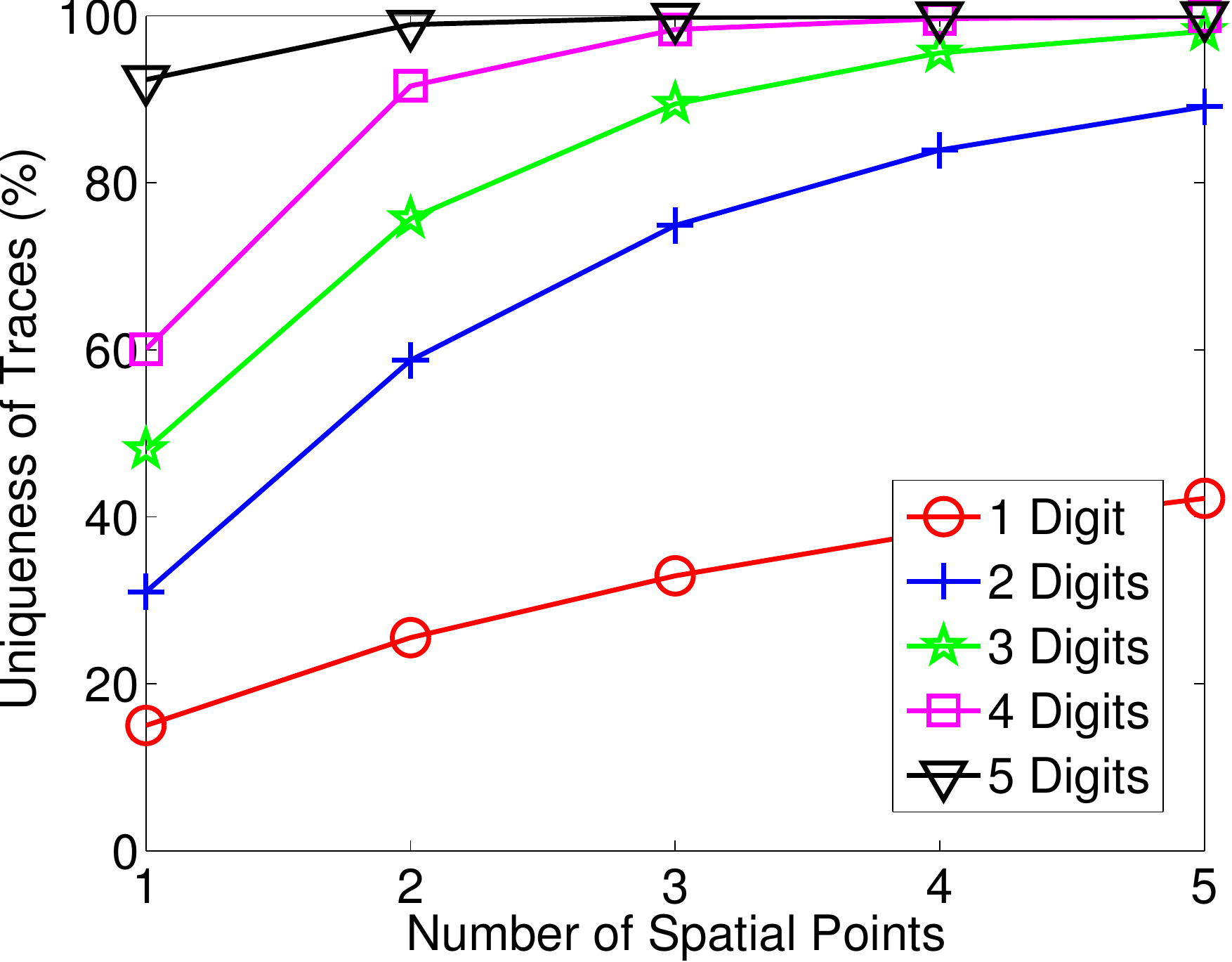}}
\hspace{0.01in}
\subfigure[GeoLife]{\includegraphics[width=0.31\linewidth]{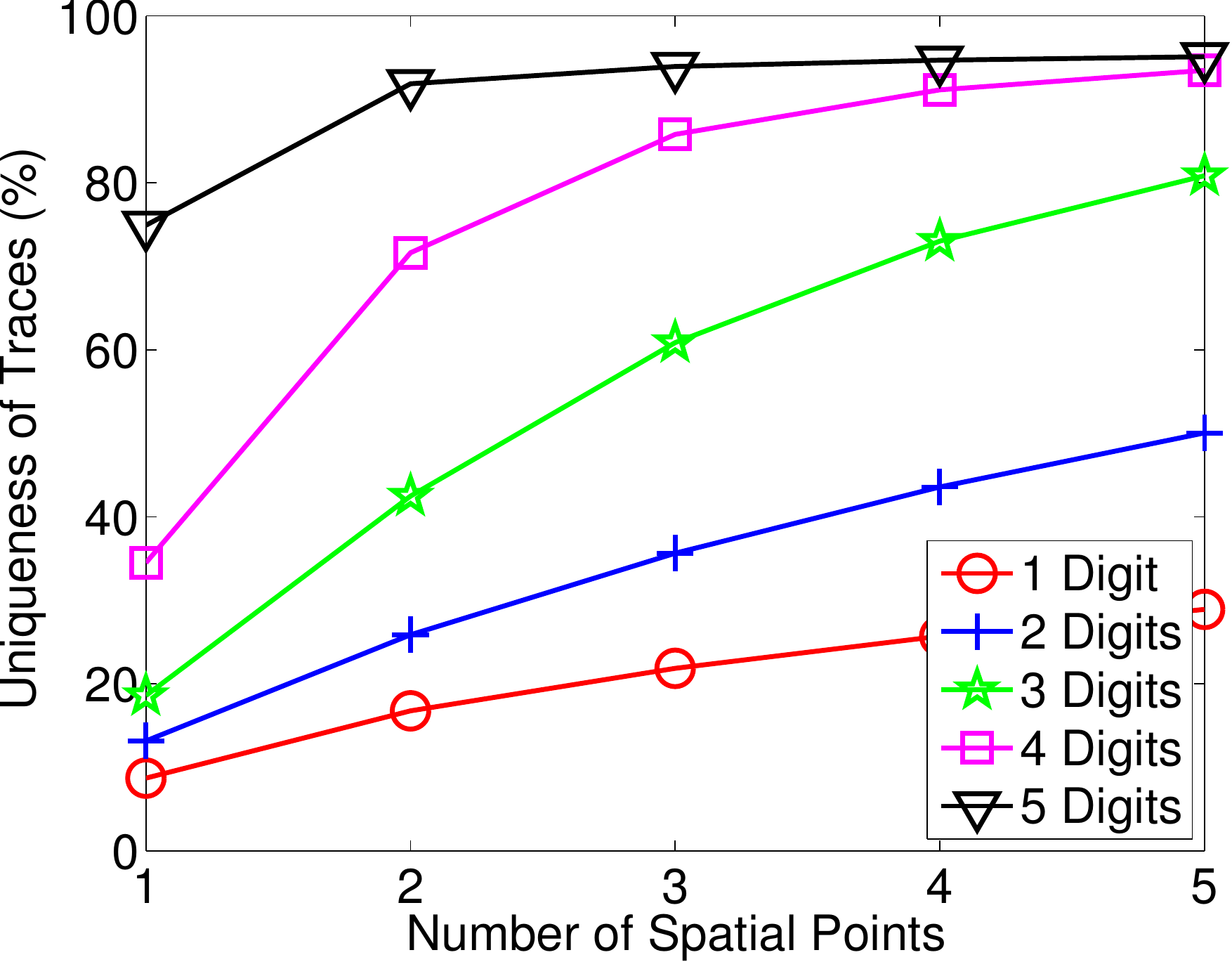}}
\caption{Average uniqueness of spatial information. For each dataset, we measure the average uniqueness of the users' location information (y-axis) as we vary the number of data points (x-axis). We consider the GPS location information at different levels of decimal resolution: 1 digit (red), 2 digits (blue), 3 digits (green), 4 digits (magenta) and 5 digits (black). Decreasing the decimal place resolution of the GPS coordinates generally leads to a lower average uniqueness, suggesting that spatial coarsening can help to obfuscate the identity of mobility data users.}
\label{resolution1}
\end{figure*}

\begin{figure*}[t!]
\centering
\subfigure[CabSpotting]{\includegraphics[width=0.46\linewidth]{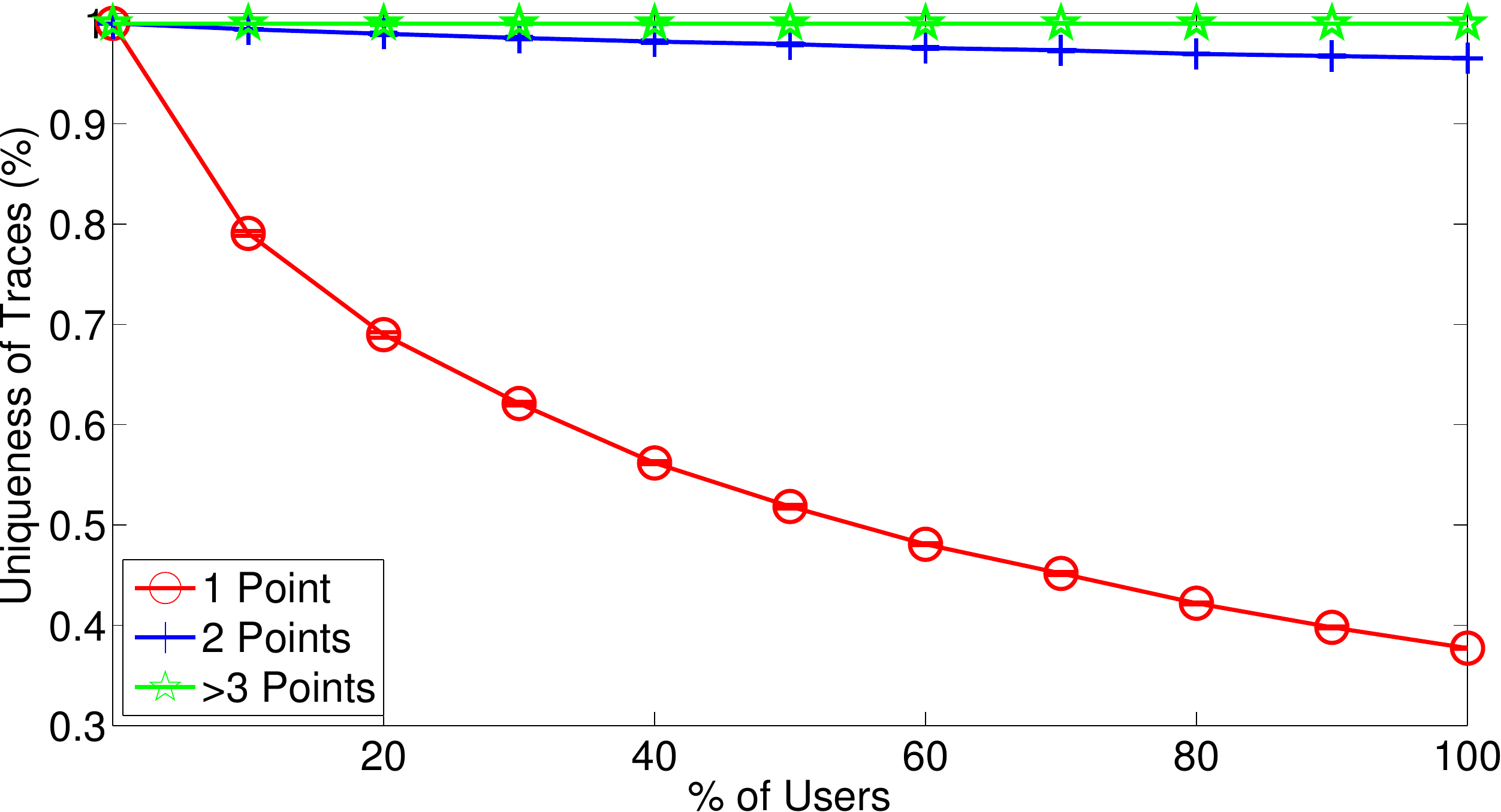}}
\hspace{0.05in}
\subfigure[GeoLife]{\includegraphics[width=0.46\linewidth]{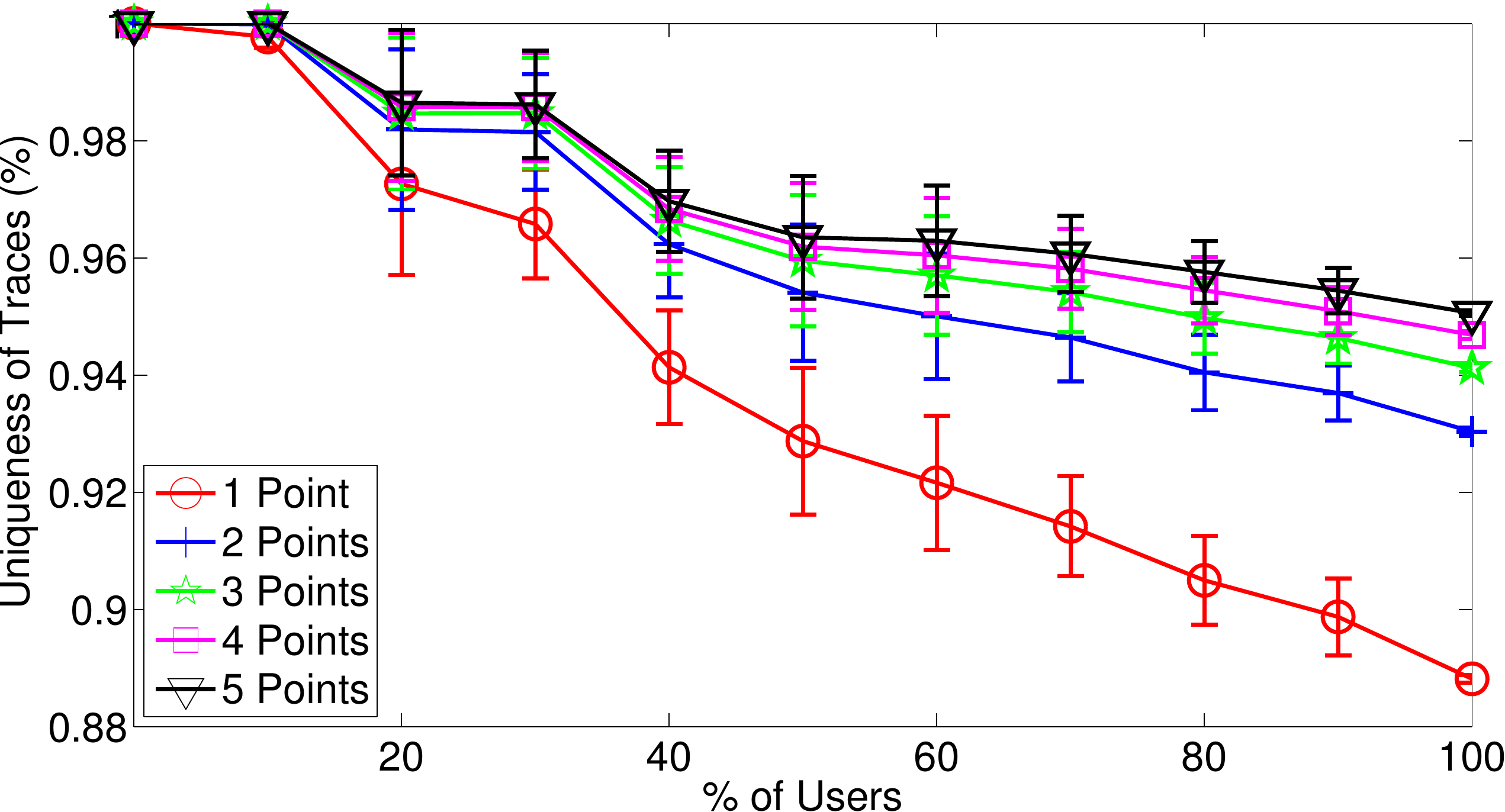}}
\caption{Effect of the number of users on the average uniqueness. We measure the average uniqueness of the users' traces (y-axis) as we vary the number of users in the datasets (x-axis). Moreover, we show how the average uniqueness varies as we change the number of spatial points considered. Note that as the number of users increases more uncertainty is added to the identity information. However, we observe that in the CabSpotting dataset the number of spatial points plays a pivotal role in the determination of the level of uniqueness.}
\label{users}
\end{figure*}

We evaluate the uniqueness of a mobility trace as follows. Given a trajectory $M$, let $S_n(M)$ be a subset of $n$ points taken from $M$. We say that $S_n(M)$ uniquely identifies the single trace $M$ when the number of traces that contain $S_n(M)$ is one. Let $m(S_n(M))$ denote the number of traces which are uniquely identified by $S_n(M)$: the lower $m(S_n(M))$ the more unique a trace is. In these experiments, for each user we sample $1000$ random subsets $S_n(M)$. We then evaluate the uniqueness of a human mobility trace as the percentage of subsets $S_n(M)$ that matches exactly one trace, i.e., $m(S_n(M))=1$. For each dataset, the results are presented in terms of average uniqueness over the whole dataset, with a $95\%$ confidence interval. The same procedure is repeated for spatial points, spatio-temporal points and the movement signatures described in the previous section. 

As a first analysis, we evaluate the uniqueness of a single spatial and spatio-temporal point. Due to the precise nature of GPS information, we expect the uniqueness of these points to be very high, provided that the spatio-temporal information is sufficiently accurate. In fact, we observe that taking the temporal dimension into account raises the uniqueness of the traces to nearly $100\%$, over 2 out of 3 datasets. In the GeoLife dataset, despite the high spatio-temporal resolution, we measure a uniqueness of $89.5\%$. We suspect that this is due to some duplicated trajectories. However, when the spatial location alone is taken into account, the uniqueness over the three datasets can be considerably different, remaining around $100\%$ for CenceMe and $89\%$ for GeoLife but dropping under $40\%$ for CabSpotting. This in turn suggests that the efficacy of anonymization methods that rely on attributes suppression to enforce k-anonimity~\cite{sweeney2002achieving} is largely dependent on the nature of the dataset. In fact, a possible reason for the high uniqueness of the mobility traces of the GeoLife and CenceMe datasets is that they contain mobility traces of users through their daily routines which, unlike the mobility of taxi cabs, contain many personal and unique locations such as home and work. In particular, a close look at the traces in the CenceMe dataset reveals that in several cases the GPS coordinates remain effectively constant for long periods, thus inevitably raising the uniqueness of that location. This may be for example the case of a person sitting in his or her office. Despite making the identification task trivial, this is of particular interest, as our daily movements patterns do include this kind of very personal and unique location.

Not only spatial and temporal points do uniquely identify users, but the characteristics of movements can also be highly individual. Fig.~\ref{fig:movement} shows the average uniqueness as we vary the number of points describing the average speed, the total distance and the average direction over a time window of 30 seconds. Note that, due to the coarser temporal resolution of the CenceMe dataset, for this dataset we use a time window of 1 hour. As we can see, the results are largely dependent on the dataset and on the number of points used. For example, on the CenceMe dataset as little as 3 points are sufficient to uniquely identify $100\%$ of the individuals. Most remarkably, we observe that in all the three datasets the average direction of travel is the most discriminative feature, while the average covered distance is the least discriminative one. In fact, with the exception of the CenceMe dataset, in the GeoLife and CabSpotting datasets the average distance is a very poor signature of a person's movements, whereas the 5 average direction points are sufficient to uniquely identify more than $95\%$ of the users.

\begin{figure*}[t!]
\centering
\subfigure[CabSpotting]{\includegraphics[width=0.317\linewidth]{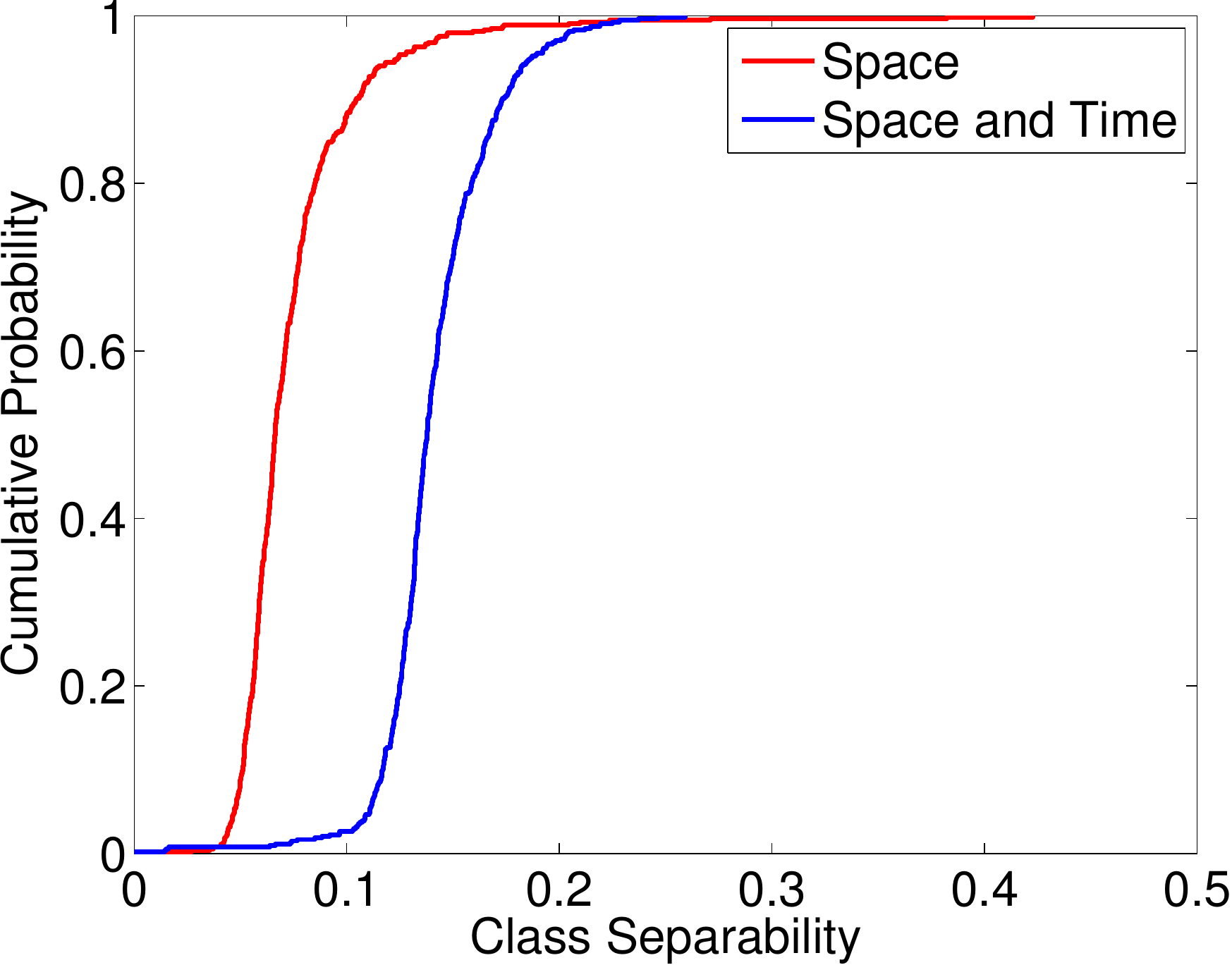}}
\hspace{0.01in}
\subfigure[CenceMe]{\includegraphics[width=0.31\linewidth]{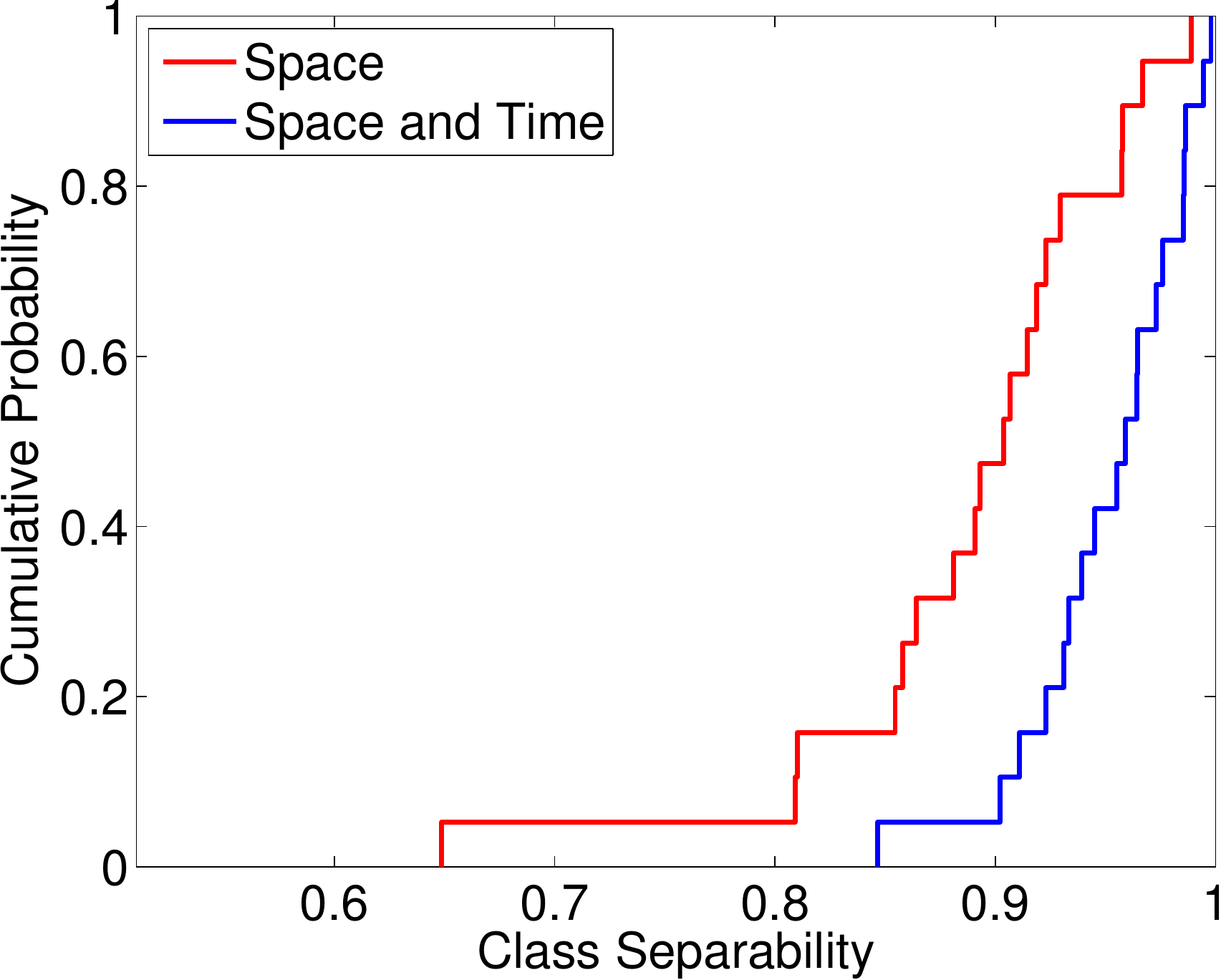}}
\hspace{0.01in}
\subfigure[GeoLife]{\includegraphics[width=0.31\linewidth]{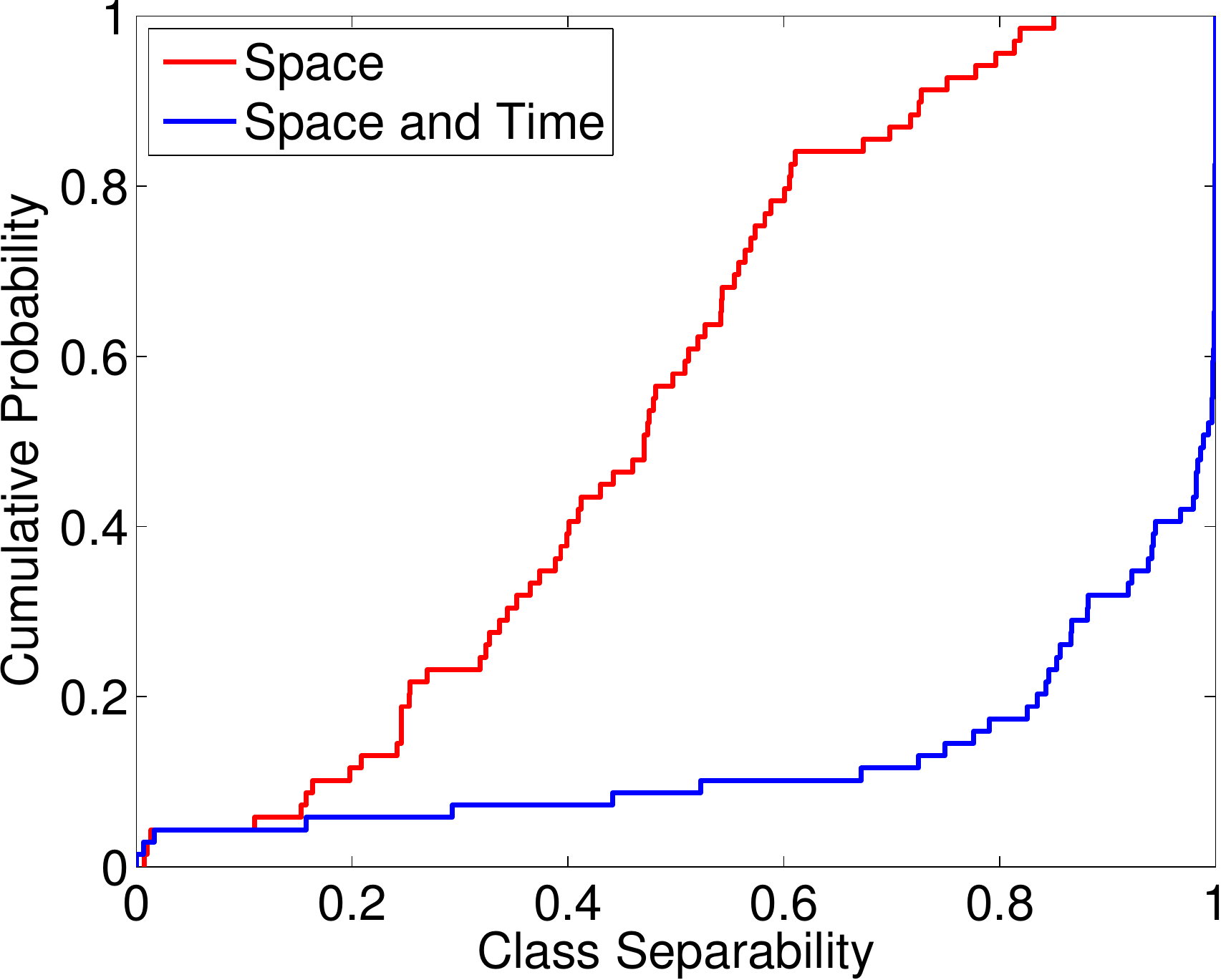}}
\caption{Empirical CDF of the geometric separability over the three datasets. For each dataset, we show the cumulative probability (y-axis) of the per class geometrical separability (x-axis). The red line refers to the case where only the spatial information is used, whereas the blue line refers to the case in which also time is taken into account. We observe that in general the addition of the temporal dimension makes the data much easier to separate.}
\label{separability}
\end{figure*}

We then consider spatial information alone, and we investigate how the uniqueness of traces varies as the number of sampled points is increased. Moreover, we show that a few points, even with reduced resolution, are enough to uniquely identify a large number of the users. We reduce the resolution of spatial points by truncating the latitude and longitude values to fewer decimal places, effectively coarsening the spatial information of the traces~\cite{sweeney2002achieving,abul2008never}. In the original CabSpotting dataset the spatial precision of points is of 5 decimal places, which represent an area of approximately 1.11 by 0.96 meters~\cite{robinson1960elements}. A 4 decimal places resolution, on the other hand, represents an area of approximately 11.09 by 9.55 meters, while a 1 decimal place resolution represents a patch as large as 11087.4 by 9550.6 meters. Hence, it is interesting to investigate to which extent this spatial coarsening can help to obfuscate the identity of mobility data users. Fig.~\ref{resolution1} shows the average uniqueness over the three datasets as the decimal place resolution and the number of sampled points vary. 

When the full-resolution 5 digits GPS coordinates are used, we have that in both the CabSpotting and the CenceMe datasets as little as two points are sufficient to uniquely identify nearly $100\%$ of the individuals. In the GeoLife dataset, on the other hand, sampling more points results in a slower increase of the uniqueness, thus suggesting the existence of a considerable spatial overlap between different traces. In fact, in all the experiments we observe a clear upper bound in the uniqueness of this dataset, which is due to some of the traces sharing the exact same series of spatio-temporal points.
Most importantly, Fig.~\ref{resolution1} also shows that a considerable number of users can be still identified by a small fraction of very coarse spatial points. However, we observe a marked drop of the average uniqueness in the CabSpotting dataset when the decimal place resolution is less than 4 digits, which once again highlights the fact that this dataset contains less unique locations than the other two datasets. Note that, from a practical point of view, these findings are of particular importance to Android users, where the location privacy permissions of applications can be set to access either coarse- or fine-grained location information\footnote{https://developer.android.com/google/play-services/location.html}.

\begin{figure*}[t!]
\centering
\subfigure[CabSpotting]{\includegraphics[width=0.31\linewidth]{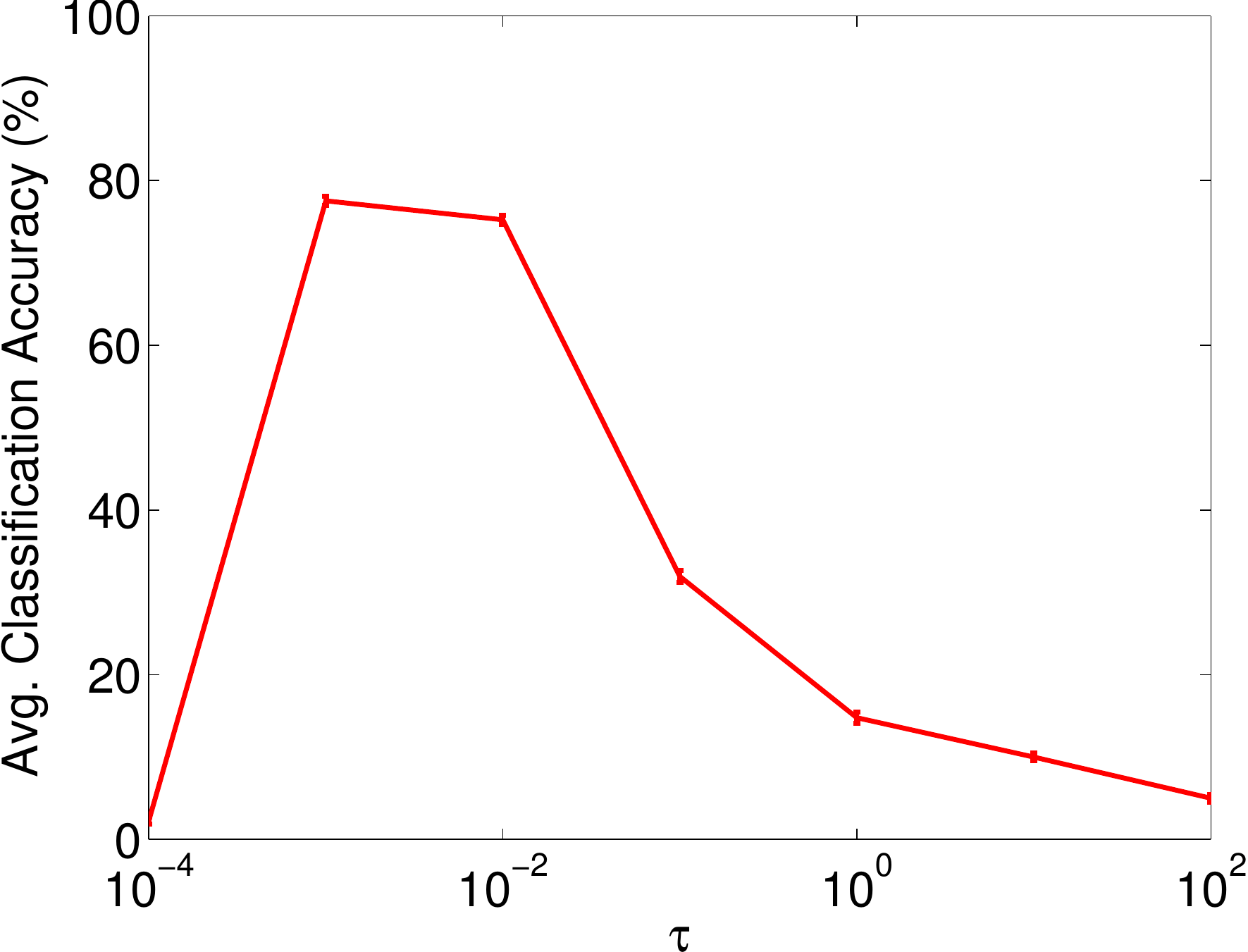}}
\hspace{0.01in}
\subfigure[CenceMe]{\includegraphics[width=0.31\linewidth]{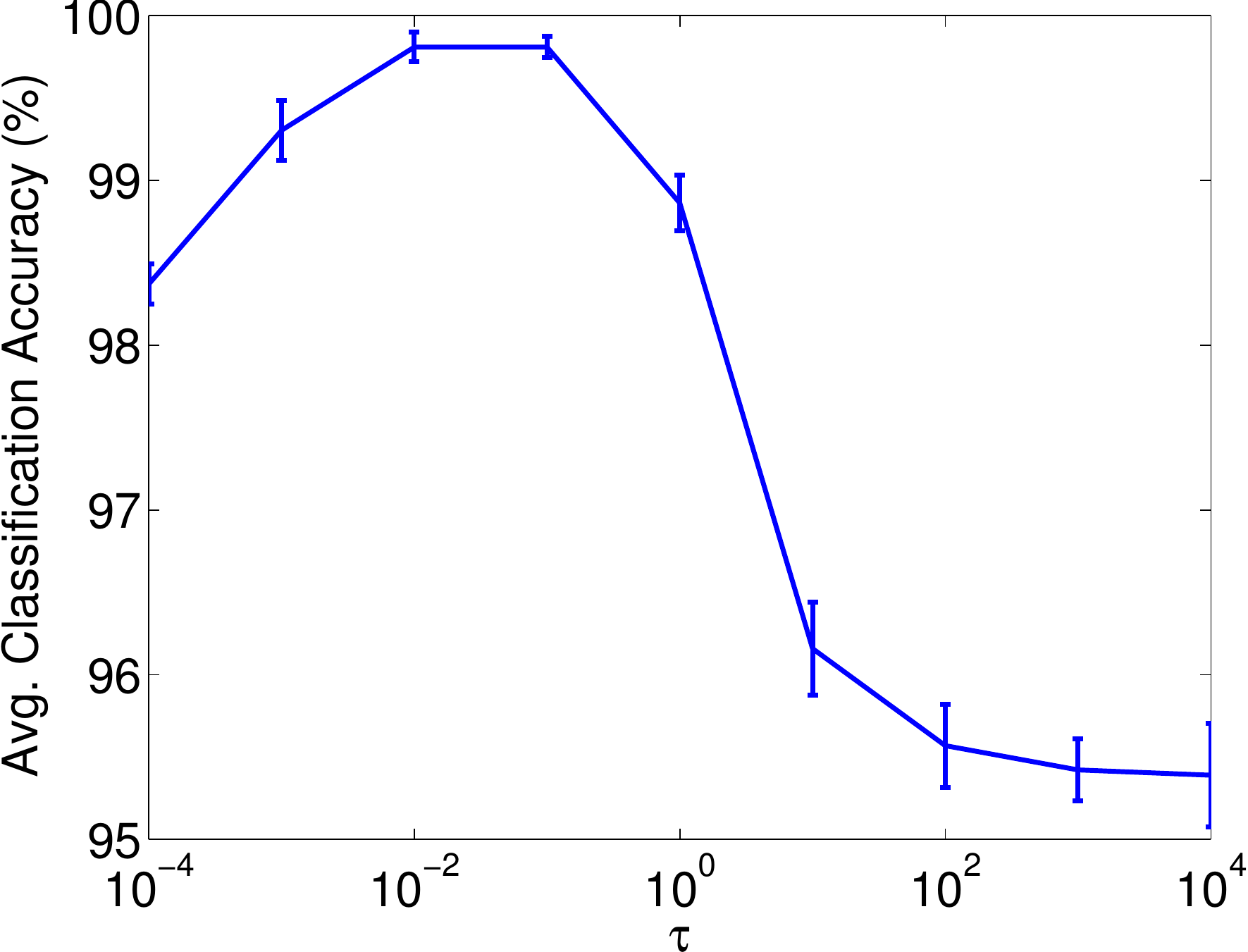}}
\hspace{0.01in}
\subfigure[GeoLife]{\includegraphics[width=0.31\linewidth]{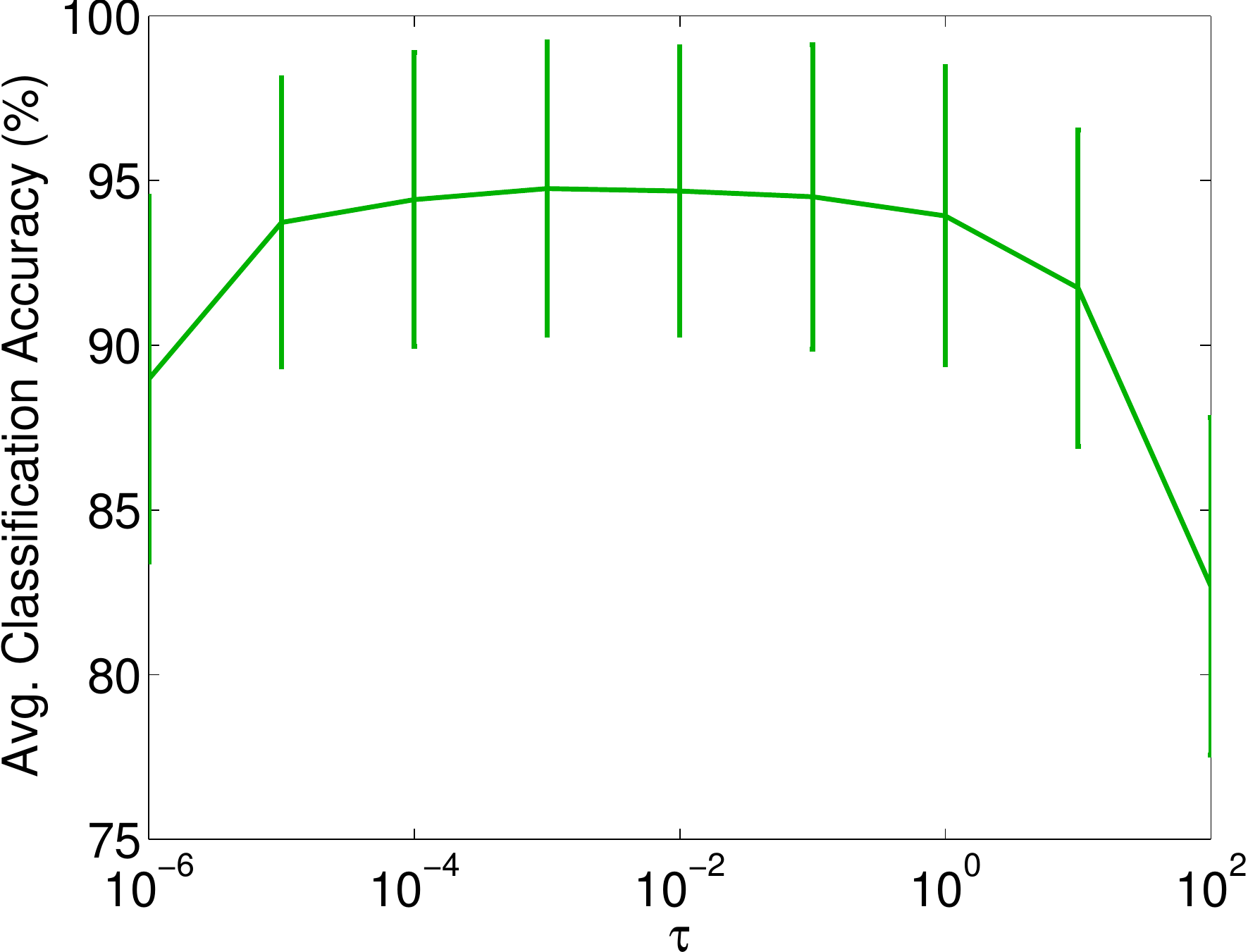}}
\caption{Optimization of the $\tau$ parameter. For each dataset, we measure the impact of the value of $\tau$ (x-axis) on the average classification accuracy (y-axis). The optimal $\tau$ is selected by maximizing the classification accuracy on 3 unseen points. Note the difference in the y-axis scale.}
\label{tau}
\end{figure*}

As a last experiment, we measure how the uniqueness varies as we increase the number of users in the datasets. Since the uniqueness of the CenceMe dataset is already near $100\%$ when the whole dataset is considered, we limit this analysis to the CabSpotting and GeoLife datasets. Fig.~\ref{users} shows the value of the average uniqueness as the number of users in the datasets increases. The limit case is that of a dataset containing a single trajectory, which has $100\%$ uniqueness as there is no uncertainty in the identity of the user. On the other hand, as the number of users increases, the uniqueness starts to decrease, since more and more uncertainty is added to the identity information. Remarkably, we see that the number of points in $P$ plays a fundamental role in the determination of the level of uniqueness in the CabSpotting dataset. Here increasing the number of points to 3 or more raises the uniqueness to $100\%$, regardless of the number of users in the dataset.

\subsection{Classification of Previously Unseen Data}
In this subsection we focus on the problem of classifying unseen points, i.e., points that in our experiments we assume not present in the datasets associated to certain individuals, by using the distance function described in the methodology section. Before this, however, we study the separation properties of the trajectories for the three dataset being investigated.

\subsubsection{Analysis of Trajectory Separability}
Before turning to the problem of classifying unseen points, we propose a simple yet effective way to quantify the difficulty of the classification problem. More specifically, we propose to measure the geometric separability of the trajectories of a given dataset~\cite{thornton1998separability}. The Geometric Separability Index was introduced by Thornton~\cite{thornton1998separability} to measure the degree to which data belonging to the same class tend to cluster together. This is done by estimating the proportion of points in the dataset whose nearest-neighbour belongs to the same class, i.e.,
\begin{equation}
GSI(f) = \frac{| \lbrace p | f(p) = f(n(p)) \rbrace |}{N},
\end{equation}
where $N$ is the number of points in the dataset, $n(p)$ denotes the nearest-neighbor of $p$ and $f$ is a binary function assigning a point to a class. Note that the GSI measures separability in a more general sense than linear separability. In fact, the data may be non-linearly separable but still geometrically separable. Consider for example two sets of points along concentric circles with different diameters, which are not linearly separable yet are clustered along clearly separate structures. In general, the GSI ranges from 0 to 1, with a value of 1 for two completely separated clusters and a value of 0 for two completely overlapping clusters.

\begin{figure*}[t!]
\centering
\subfigure[CabSpotting]{\includegraphics[width=0.31\linewidth]{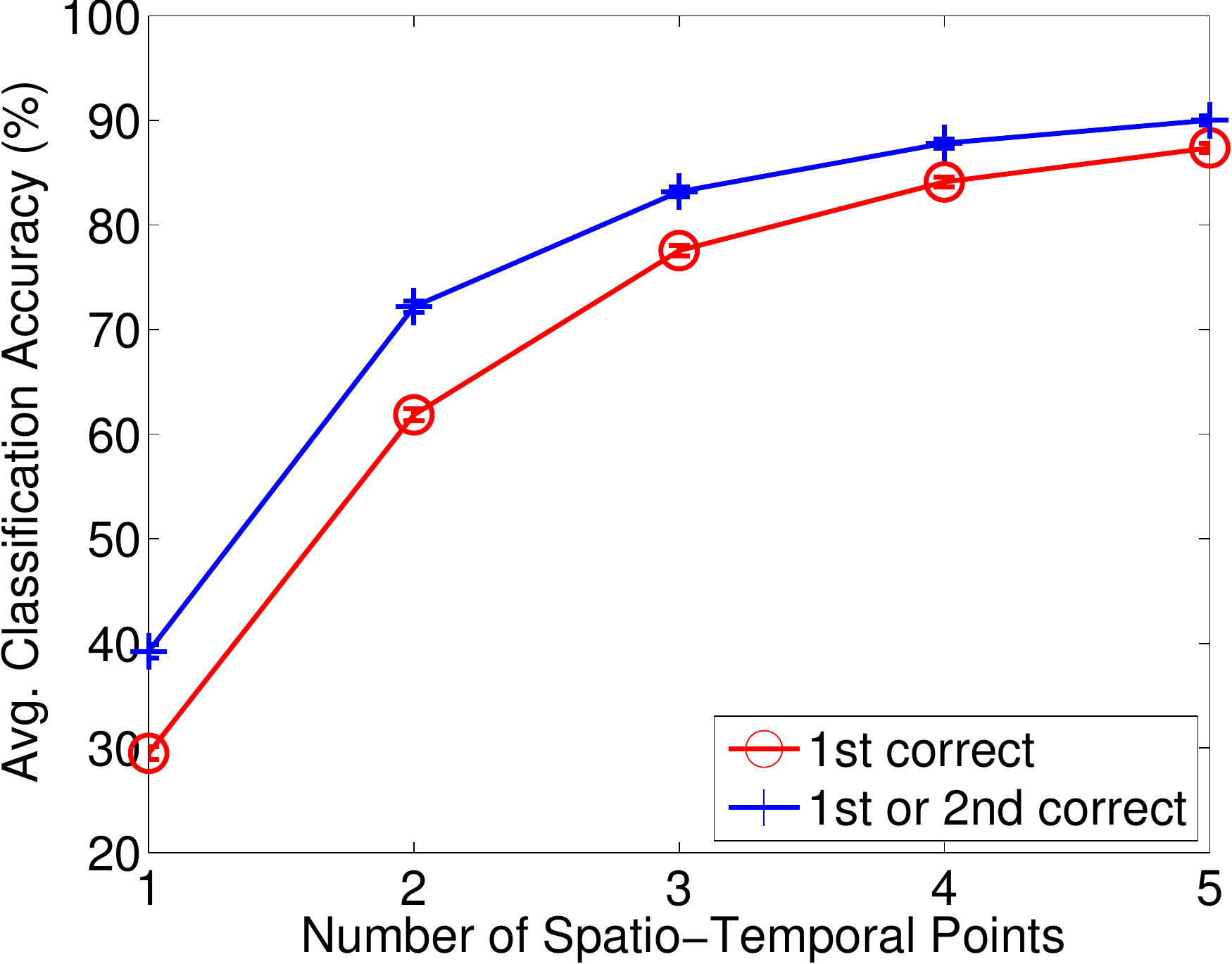}}
\subfigure[CenceMe]{\includegraphics[width=0.31\linewidth]{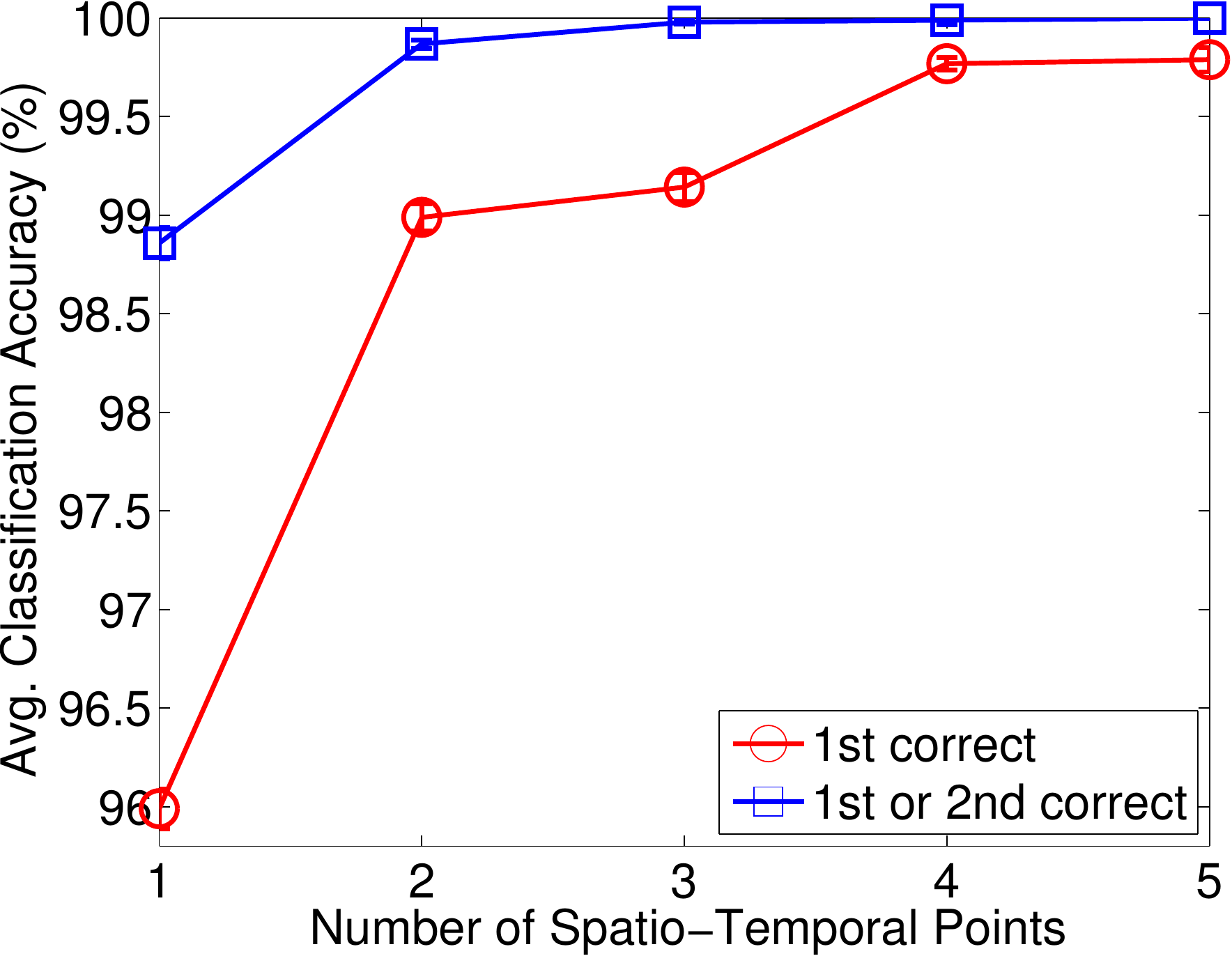}}
\subfigure[GeoLife]{\includegraphics[width=0.31\linewidth]{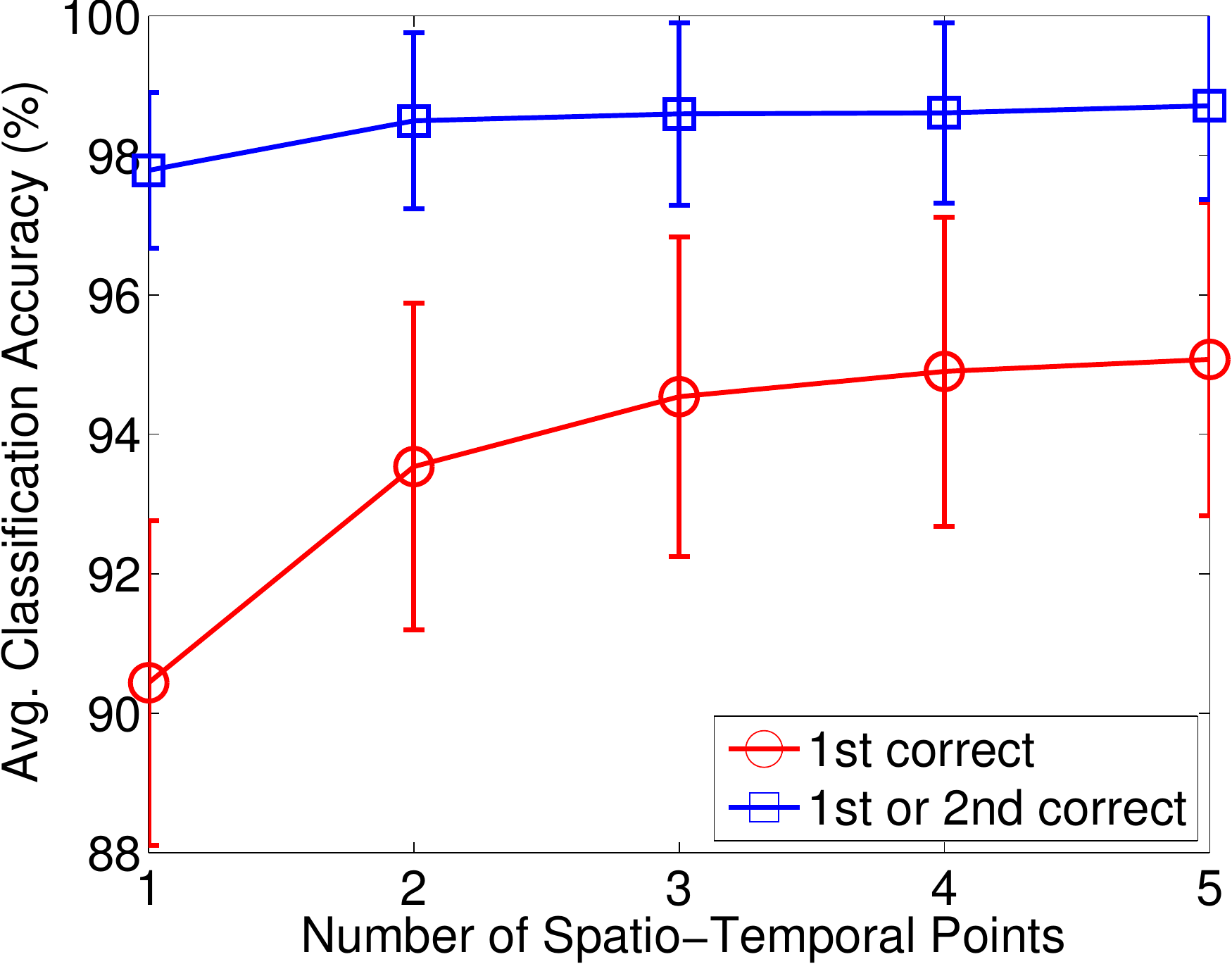}}
\caption{Average classification accuracy over the three datasets. For each dataset, we show the average classification accuracy (y-axis) as we increase the number of sampled spatio-temporal points (x-axis). Note that as little as 1 spatio-temporal point is sufficient to correctly classify more than 90\% of the users, in 2 out of 3 datasets. Here the red line refers to the case where the correct trajectory is the nearest one, whereas the red line refers to the case where the correct trajectory is either the first or the second nearest one.}
\label{unseen}
\end{figure*}

\begin{figure}[!t]
\centering
\includegraphics[width=1\linewidth]{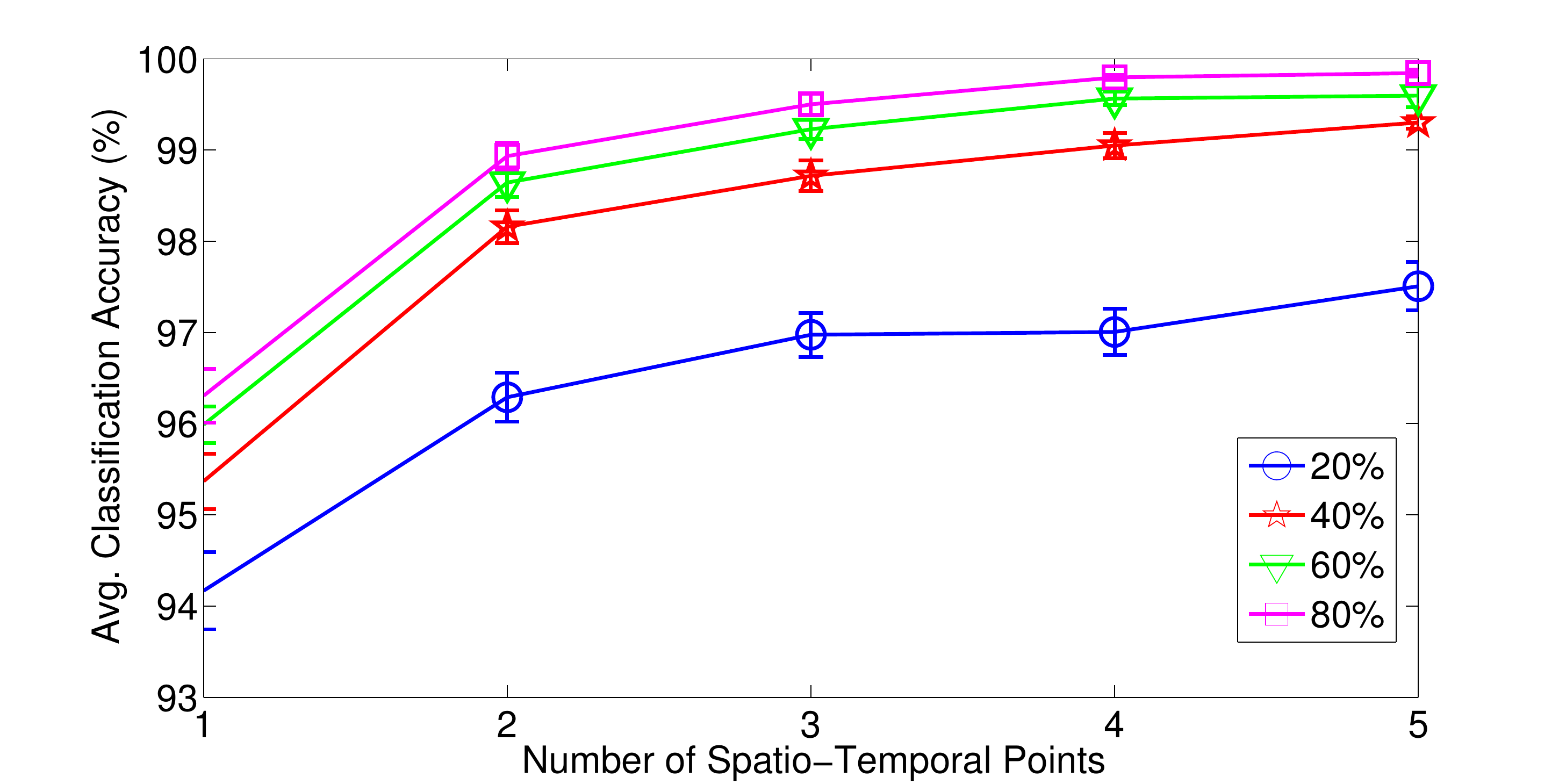}
\caption{Effect of training set size on the classification accuracy. We measure the average classification accuracy (y-axis) as we increase the number of sampled spatio-temporal points (x-axis) for the CenceMe dataset. We show how the results vary as we reduce the size of the observed traces in the CenceMe dataset to 20\% (blue), 40\% (red), 60\% (green) and 80\% (magenta) of the original size. The results show that our classification framework is robust with respect to the number of available observations.}
\label{train}
\end{figure}

The GSI was originally introduced to measure the separability of data in a binary classification problem, i.e., where the number of possible classes is two, while here we are dealing with a number of classes equal to the number of users in the dataset. Moreover, we can see that the GSI does not take the size of the different classes into account. That is, a large and easily separable class will impact the GSI much more than another small and non-separable class. Here we propose to take the average GSI (aGSI) over all the classes, which we define as
\begin{equation}
aGSI(f) = \frac{1}{N}\sum_{c \in C}\frac{| \lbrace p_c | f(p_c) = f(n(p_c)) \rbrace |}{N_C},
\end{equation}
where $f$ is a function that assigns a class to a point, $C$ is the set of classes, $p_c$ denotes a point belonging to $c \in C$ and $N_C$ denotes the number of such points.

We find that the aGSI of the CabSpotting dataset is $0.0741$, the aGSI of the CenceMe dataset is $0.8438$ and the aGSI of the GeoLife dataset is $0.4437$. This confirms our observation that the CabSpotting dataset is the most challenging one. Fig.~\ref{separability} shows the empirical cumulative distribution function of the per class geometrical separability over the three datasets. Here the red line shows the empirical CDF when only the spatial information is used, while the blue line refers to the case in which also time is taken into account. Note in particular that in the GeoLife dataset the addition of the temporal dimension makes the data much easier to separate. Moreover, Fig.~\ref{separability} highlights once again that the CabSpotting dataset is the most challenging one in that the different trajectories show a high degree of overlap.
\subsubsection{Classification Results}
We now turn to the problem of classifying unseen points. Here, we do not analyze the uniqueness of a subset of points from a mobility trace, but instead we analyze the similarities between a set of given points, which we refer to as sampled points, and a set of disjoint mobility traces. The similarity is then used to identify a person associating him/her with the nearest mobility traces using Eq.~\ref{final_distance}. Note that we are working under the assumption that the sampled points all belong to a single mobility trace. For each trace in the dataset we sample a set of $n$ points and we compute the nearest and the second nearest neighbor among the labeled traces, where the above points have been removed. We repeat this procedure $100$ times to compute the average classification accuracy, with a $95\%$ confidence interval. For each dataset, the results are presented in terms of average accuracy over the whole dataset. Recall that in these datasets each person is mapped to a single trajectory. Hence, the accuracy is measured as the number of individuals that are correctly matched to their trajectory, where an individual is represented by the set of $n$ sampled points defined above.

Recall that according to Eq.~\ref{distance} the distance between two spatio-temporal points $p_1$ and $p_2$ is computed as $d_s(p_1,p_2) e^{\frac{d_t(p_1,p_2)}{\tau}}$, where $\tau$ controls the effect of the temporal difference between the two points. In other words, the larger the value of $\tau$, the less relevant is the temporal difference between the points into account. As a consequence, the distance between a set of spatio-temporal points and a mobility trace (Eq.~\ref{final_distance}) also depends on the choice of the parameter $\tau$.
We randomly partition each dataset into a training and test set, where each trace contains $50\%$ of the original GPS points. Given the training set of traces, we calculate the optimal order of magnitude of $\tau$ in terms of average classification accuracy of 3 test points. Fig.~\ref{tau} shows the value of the average classification accuracy for increasing values of $\tau$, with a $95\%$ confidence interval. We observe that in all the three datasets the optimal value of $\tau$ varies between $10^{-2}$ and $10^{-3}$. This confirms the intuition that the temporal dimension has to be taken into account to yield a higher classification accuracy. Indeed, choosing a small value of $\tau$ amounts to restricting the focus of our nearest neighbor search to those points that lie both spatially and temporally close to that being classified, which are indeed more likely to belong to the correct trace. Also, note that in the CabSpotting dataset the accuracy of our method is particularly influenced by the value of $\tau$, and thus it is critical to properly optimize it before proceeding to the classification phase. On the other hand, we observe that the accuracy on the GeoLife dataset remains essentially constant between $10^{-4}$ and $10^{-1}$. This may be partially due to the high temporal density of the trajectories. In $91\%$ of the trajectories, in fact, the GPS location was sampled every $1 \sim 5$ seconds. As a result of this, the high resolution temporal information (i.e., low $\tau$) can be used to discriminate between the different trajectories. This is in stark contrast with the CabSpotting dataset, where setting the value of $\tau$ to low values results in a sudden drop of performance.

After obtaining the optimal values of $\tau$, we use them in order to evaluate our classification framework on the three datasets. Fig.~\ref{unseen} shows how the average classification accuracy varies as the number of sampled points increases. We also consider the situation in which the correct label is that of the second nearest neighbor. As shown in Fig.~\ref{unseen}, as little as 1 spatio-temporal point is sufficient to correctly classify more than $90\%$ of the users, in 2 out of 3 datasets. Moreover, when the second nearest neighbor is taken into account, on the same datasets the average classification accuracy approaches $100\%$. Once again, the CabSpotting dataset proves to be the hardest to analyze given its inherent characteristics, such as common routes of taxis and the presence of locations associated to taxi ranks. Instead, the results related to the GeoLife and CenceMe datasets imply that it is possible to correctly identify the user with a very small number of new observations not present in the original one. In other words, it is harder to potentially ensure the location privacy of the individuals. We conjecture that this is mainly due to the presence of many personal and thus unique locations, such as home and workplace locations, as opposed to CabSpotting.

Finally, we evaluate the impact of the size of the traces on the classification accuracy. In particular, we reduce the size of the observed traces in the CenceMe dataset to $20\%$, $40\%$, $60\%$ and $80\%$ of the original size. We present the result of this experiment on the CenceMe dataset. 
Fig.~\ref{train} shows how the average classification accuracy varies as we increase the number of the sampled points, for different sizes of the observed traces. As expected, reducing the number of observable spatio-temporal points has the effect of lowering the average classification accuracy. However, we note that our approach still performs considerably well when as little as $20\%$ of each trace is considered, thus suggesting a good robustness against the lack of available observations.

\vspace{-0.1in}
\section{Discussion}\label{discussion}
The results presented in the previous section about the experimental validation of our approach show that users can be identified using a few high resolution location points. We believe that our findings raise important privacy concerns with respect to the management, storage and analysis of personal mobility data. Even a single high precision spatio-temporal point should be treated with great care in order to preserve the privacy of a user if additional information is available. 
More precisely, we have shown that in the datasets considered in this study a single spatio-temporal point is sufficient to uniquely identify nearly $100\%$ of the individuals. When only spatial information is considered, we find that in some cases the uniqueness of the mobility data can still get close to $100\%$. In general, a limited number of spatial points is required to uniquely identify individuals, without having the points to be classified as part of the given mobility traces used for training. Even hiding the temporal information and coarsening the spatial resolution of the points may still not be sufficient to ensure the privacy of users. 

We have also showed that the movement characteristics of speed, direction and distance of travel are very sensitive from a user privacy perspective: access to a user record of any of these movement characteristics, e.g., compass recordings, should be handled with a similar level of privacy as precise positioning data. Mobility points which have been removed from a trace should also be treated with great care. Our study of previously unseen points showed that, in the majority of cases, given a limited number of GPS spatio-temporal points it is possible to identify the traces from which the points originated, thus allowing a potential attacker to transfer the identity information from a non-anonymized dataset of trajectories to an anonymized set of points.

The ability to identify individuals from previously seen points depends on the uniqueness of the dataset of mobility traces taken into consideration. At the same time the ability to identify individuals from unseen points depends on both the uniqueness and the possibility of associating a user to a finite number of significant places or areas. Therefore, we expect that the ability to identify individuals who spend a large amount of time in the same locations should be generally low. However, in this paper we have successfully tested our identification technique on a dataset of taxi traces, which, due to the fact that the trajectories are spatially constrained to lie on the streets, is characterized by a high number of common locations.

\section{Related Work}\label{related}

In the recent years, due to the increasing popularity of mobile phones (now mostly equipped with GPS receivers), there has been a strong interest in the analysis of location privacy risks and potential countermeasures.
With respect to the problem of identification of people from human mobility, Gruteser et al.~\cite{gruteser2005anonymity} exploited the associativity of GPS mobility traces to identify individual traces from a collection of unlabeled traces of multiple users. This unmixing of traces is carried out by identifying different paths in the dataset, under the assumption that a user is likely to continue traveling along the same route. It is worth noting that their system is prone to misclassification when paths cross, as it is unable to infer whether the paths of two or more individuals actually crossed or just touched.
Recently, de Montjoye et al. have shown how unique the location of users are when they make or receive mobile phone calls or text messages~\cite{de2013unique}, or when they perform a credit card transaction~\cite{de-montjoye:reidentifiability}. In~\cite{de2013unique}, each time a user phone call is started or a text is sent or received, the location of the nearest network service antenna along with the time is recorded.  In~\cite{de-montjoye:reidentifiability}, on the other hand, each location point is a triple containing the shop identifier, time and price of a credit card transaction. Both works analyzed these location traces to find how many spatio-temporal points are needed to uniquely identify the user. The main difference with respect to these works is that in our study we do not include the points used for the classification in the training set, i.e., our work also focuses on points that are not present in the training set. Moreover, de Montjoye et al. consider traces extracted from call data records or shops visits, whereas in this work we consider GPS data points.

 
Golle and Partridge~\cite{golle2009anonymity} have shown how the uniqueness of home/work pairs can be used to carry out inference attacks to reveal the identity of a user from an anonymized trace. In addition to this, Ranjan et al.~\cite{ranjan2012call} have recently examined the use of mobile phone call records for studies in human mobility and concluded that they can be very biased to home and work locations. 
Another related study is presented in~\cite{de2008identification}, where mobile users are identified given a set of locations collected from mobile phone GSM Call Data Records. However, the locations in these recordings usually correspond to areas where a user made a mobile phone call or text, i.e., locations which are more likely to correspond to the home and workplace of a user, which are again inherently unique locations. Rossi and Musolesi recently proposed a trajectory-based and a frequency-based attack against Location-based Social Network (LBSN) users~\cite{rossi2014s}. Note, however, that in the context of LBSNs a location point is a \emph{check-in} at a venue. Check-ins are generally very sparse in space and time, and thus very different from the traces that we considered in this study. Monreale et al.~\cite{monreale2014privacy}, on the other hand, proposed to adopt the privacy-by-design paradigm in big data analytics to reach a trade-off between data privacy and quality of the data. For the case of GPS mobility traces, they propose to use a Voronoi tessellation to ensure $k$-anonymity while maximizing the quality of the anonymized data.

Note that a number of works in the literature have studied the problem of location prediction and the related privacy implications~\cite{ashbrook2003using,xue2013destination}. However, while in the case of location prediction one is interested in preventing a potential attacker to infer the next place visited by an individual, in our paper we focus on the problem of identifying the individual himself/herself, rather than the location. Given a location prediction model one can also infer the identity of a user by means of maximum likelihood estimation. However, the aim of this paper is to show that extremely accurate yet less elaborate and computationally demanding techniques can be used to disclose a user's identity. Finally, another related work is~\cite{krumm2007inference}, in which the author studies the more specific problem of inferring the home location of a user participating in a database of GPS traces. Given the estimated coordinates of a user's home, a simple Web-based lookup is used to reveal his/her name. The focus of this work is different: in fact, we are interested in evaluating the uniqueness of GPS information and the extent to which an unseen set of points can be linked to the underlying generating trajectory, thus revealing the identity of a user. 

%


\section{Conclusions}\label{conclusion}

In this paper we have introduced a series of techniques for the analysis and identification of GPS mobility traces.
We have firstly showed that it is possible to identify users with great accuracy using movement data such as speed, direction and distance of travel. Secondly, we have evaluated the uniqueness of GPS mobility traces using three popular datasets. We have analyzed the use of both spatial as well as spatio-temporal information to perform this task and we have showed that, in the datasets taken into consideration, as little as two spatial points are sufficient to uniquely identify nearly $100\%$ of the users. We have also evaluated the impact of the dataset size and the precision of the GPS coordinates on the uniqueness of the data, and we have found that, in some datasets, coarsening the GPS precision results in a drastic reduction of the average uniqueness. Finally, we have introduced a simple yet efficient technique for the identification of users from location data that are not included in the original datasets used for the training. In an attempt to quantify the extent to which a dataset can resist an identification attack like the one proposed in this paper, we have proposed a simple yet efficient way to estimate the separability of the trajectories of a given dataset. Finally, we have showed the effectiveness of the proposed identification attack on the selected datasets.

We believe that these results raise important privacy concerns with respect to the treatment of personal mobility data.
Future work will investigate the possibility of exploiting additional mobility and movement information including WiFi hotspot access and smartphone sensor readings. We also plan to analyze more refined obfuscation techniques for preserving user privacy based on the findings of this work.

\section*{Acknowledgements}

This work was supported through the EPSRC Grant ``The Uncertainty of Identity: Linking Spatiotemporal Information Between Virtual and Real Worlds'' (EP/J005266/1) and ``Trajectories of Depression: Investigating the Correlation between Human Mobility Patterns and Mental Health Problems by means of Smartphones'' (EP/L006340/1).

\bibliographystyle{unsrt} 
\bibliography{biblio}

\end{document}